\documentclass[symmetry,article,accept,moreauthors,pdftex,a4paper]{mdpi} 

\DeclareMathOperator{\sech}{sech}
\firstpage{1} 
\pubvolume{11}
\issuenum{2}
\articlenumber{271}
\pubyear{2019}
\copyrightyear{2019}
\history{Received: 9 January 2019; Accepted: 12 February 2019; Published: 20 February 2019}
\updates{yes} 

\usepackage{subfigure}
\makeatletter
\renewcommand{\@thesubfigure}{\normalsize(\textbf{\alph{subfigure}})}
\makeatother 

\Title{Soliton and Breather Splitting on Star Graphs from Tricrystal Josephson Junctions}

\Author{Hadi Susanto $^{1}$\orcidA{}, Natanael Karjanto $^{2,}$*\orcidB{},  Zulkarnain $^{3}$, Toto Nusantara $^{4}$\orcidD{} \\ and Taufiq Widjanarko $^{5}$\orcidE{}}

\AuthorNames{Hadi Susanto, Natanael Karjanto,  Zulkarnain, Toto Nusantara and Taufiq Widjanarko}

\address{%
$^{1}$ \quad Department of Mathematical Sciences, University of Essex, Colchester CO4 3SQ, UK; hsusanto@essex.ac.uk\\
$^{2}$ \quad Department of Mathematics, University College, 
Sungkyunkwan University, Natural Science Campus, 2066~Seobu-ro, Jangan-gu, Suwon 16419, Gyeonggi-do, Korea\\
$^{3}$ \quad Department of Mathematics, Faculty of Mathematics and Natural Sciences, Universitas Riau, Kampus Bina~Widya KM 12.5, Simpang Baru, Tampan, Kota Pekanbaru, Riau 28293, Indonesia; zulqr27@gmail.com\\
$^{4}$ \quad Department of Mathematics, Faculty of Mathematics and Natural Sciences, Malang State University, Jalan Semarang 5, Malang~65145,~Indonesia; toto.nusantara.fmipa@um.ac.id\\
$^{5}$ \quad Manufacturing Metrology Team, Faculty of Engineering, The~University of Nottingham, Advanced~Manufacturing Building, Jubilee Campus, Wollaton Road, Nottingham NG8 1BB, UK; Taufiq.Widjanarko@nottingham.ac.uk}

\corres{Correspondence: natanael@skku.edu}

\abstract{We consider the interactions of traveling localized wave solutions with a vertex in a star graph domain that describes multiple Josephson junctions with a common/branch point (i.e., tricrystal junctions). The~system is modeled by the sine-Gordon equation. The~vertex is represented by boundary conditions that are determined by the continuity of the magnetic field and vanishing total fluxes. When one considers small-amplitude breather solutions, the system can be reduced into the nonlinear Schr\"odinger equation posed on a star graph. Using the equation, we show that a high-velocity incoming soliton is split into a transmitted component and a reflected one. The~transmission is shown to be in good agreement with the transmission rate of plane waves in the linear Schr\"odinger equation on the same graph (i.e., a quantum graph). In~the context of the sine-Gordon equation, small-amplitude breathers show similar qualitative behaviors, while large-amplitude ones produce complex dynamics.} 


\begin{document}

\section{Introduction}

A quantum graph is a metric graph, i.e., a network-shaped structure of vertices connected by edges, with a Schr\"odinger-like operator suitably defined on functions that are supported on the edges. It arises as a model for wave propagations in a system similar to a thin neighborhood of a graph. Pauling~\cite{pauling39}  was most likely the pioneer of the research subject when he modeled free electrons in organic molecules. In~his model, he approximated the atoms as vertices while the electrons form bonds that fix a frame in the shape of the molecule on which the free electrons are confined. The~term `quantum graph' itself may be a shortening of the title of a paper by Kottos and Smilansky~\cite{kott97}. See,~e.g.,~\cite{berko16} for an elementary introduction to quantum graphs, where some basic tools in the spectral theory of the Schr\"odinger operator on metric graphs are discussed. 

Quantum graphs have been used to describe a variety of mathematical concepts as well as physical problems and applications. A review of quantum graphs with applications in theoretical physics is provided by Gnutzmann and Smilansky~\cite{gnut06}. For a comprehensive introduction and survey of the current state of research  on quantum graphs and their applications, the reader is encouraged to consult a mathematically oriented book by Berkolaiko and Kuchment~\cite{berko13}. See also an introduction and a brief survey to quantum graphs by Kuchment~\cite{kuchment08}.

The study of nonlinear counterparts of quantum graphs, where the linear wave equations are  replaced by nonlinear ones, has been growing due to their potential of becoming a paradigm model for topological effects in nonlinear wave propagation (see~\cite{noja14} for a recent review). Because of the nonlinearity, soliton solutions exist. However, a unique 'trapped soliton' state, which is admitted by the cubic focusing nonlinear Schr\"odinger (NLS) equation on the star graph with Kirchhoff conditions at the vertex, is not the ground state~\cite{adami12a}. This is remarkably different from the NLS equation on the line. The~ existence and behavior of trapped solitons with a $\delta$-interaction at the vertex are considered by Adami et al.~\cite{adami12b}. A generalized NLS equation with power nonlinearity on star graphs has also been investigated in various reports~\cite{adami14,adami16,kairzhan17}. The~existence of ground states of the same equation on several types of star metric graphs has been considered in~\cite{li18}. Bifurcations of stationary solutions in various other simple topologies have also been studied, such as in tadpole graphs consisting of a half-line joined to a loop at a single vertex~\cite{noja15}, dumbbell-shaped metric graphs~\cite{marz16,good17}, bowtie graphs~\cite{good17}, and double-bridge graphs~\cite{noja19}.

In addition to stationary solutions, the interaction of a moving soliton with the vertex is also intriguing. Soliton scattering in the NLS equation on a star graph with a repulsive $\delta$-, $\delta'$-function potential, and the free Kirchoff condition at the vertex is studied by Adami et al.~\cite{adami11}, who extended the work of Holmer et al.~\cite{holm07}. Adami et al. showed that a soliton will split into a transmitted soliton, a reflected one, and some radiation upon collision with the vertex. Soliton dynamics in star graphs with 'integrable' vertex conditions that preserve the solution norm was studied in~\cite{sobi10}.

In this work, we consider a star graph that models a tricrystal Josephson junction, see Figure~\ref{sketch}. A~Josephson junction is a quantum mechanical structure that is made of two superconducting electrodes separated by a thin barrier. Three semi-infinite junctions with the ends meeting at a common point form a tricrystal junction. The~vertex conditions were likely first derived in~\cite{naka76,naka78}, where the structure was proposed as a logic gate device. In~recent work, tricrystal Josephson junctions were fabricated as a probe of the order parameter symmetry of high-temperature superconductors~\cite{tsue94,mill95,tsue00a,tsue00}. 
In particular, tetracrystal junctions (i.e., star graphs with four arms) were also constructed and studied experimentally, see, e.g., Section IV.C of~\cite{tsue00,toma12}.

The study of soliton solutions in tricrystal junctions has only been done for topological solitons, i.e., kinks, especially when they are static~\cite{tsue00a,koga00,susa04,susa05}. This type of solitons is also called `fluxons' because they carry integer quanta of electromagnetic flux. The~dynamics of moving vortices in tricrystal junctions was discussed in~\cite{naka76,naka78,grun93,hatt96}. Here, instead, we consider for the first time the dynamics of non-topological solitons, i.e., breathers (see Figure~\ref{sketch}). In~the case of small-amplitude breather solutions, the governing equation, which is the sine-Gordon (sG) equation, can be reduced into the NLS equation with vertex conditions different from those considered in previous works. Soliton dynamics within that approximation will be discussed as well. 

The paper is structured as follows. In~Section \ref{sec2}, we discuss the governing equation of tricrystal Josephson junctions. Under an assumption of small-amplitude solutions, we will also derive the NLS equation with vertex conditions. In~Section \ref{sec3}, the scattering of linear plane waves, soliton and breather solutions  will be discussed. Numerical simulations will be presented in the same section describing the scattering processes when nonlinearity is present. In~Section \ref{sec4}, we consider the nonlinear scattering of sG breathers numerically. We present two different typical cases corresponding to small- and large-amplitudes with slow- and fast-incoming velocities. The~conclusion of the paper is in Section \ref{sec5}.

\section{Governing Equations}			\label{sec2}

\begin{figure}[htbp]
\centering 
\includegraphics[scale=0.5]{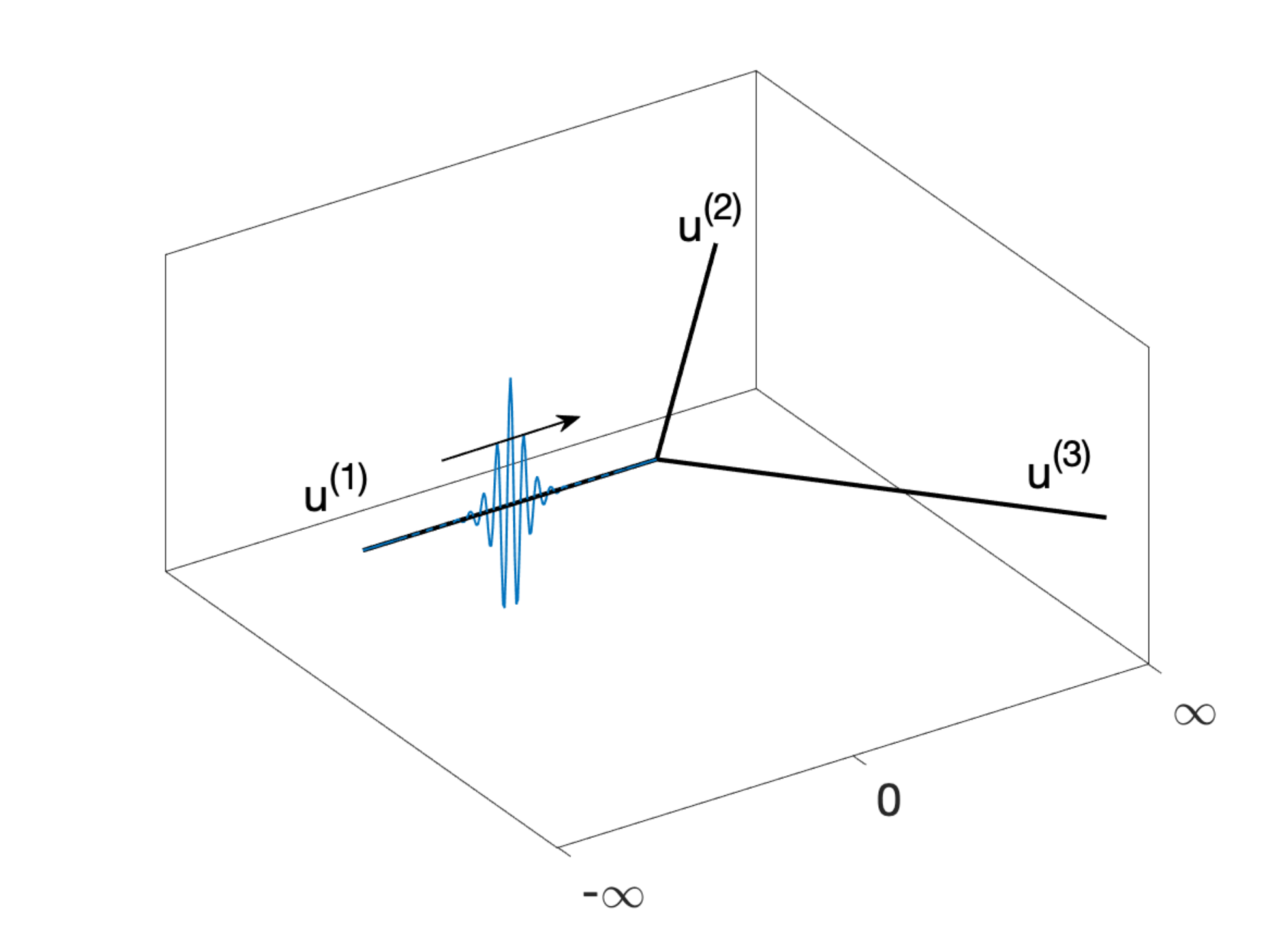}
\caption{A schematic diagram of the system, showing a breather traveling from $x\ll0$ towards $x=0$. The~convention of the spatial direction used herein is indicated.}
\label{sketch}
\end{figure}

The phase difference of wave functions along each Josephson junction is described by the sG~equation
\begin{equation}
u^{(j)}_{xx}-u^{(j)}_{tt}=\sin u^{(j)},
\label{sg}
\end{equation}
where upper indices $j=1,2,3,$ label the different branches of the system and the subscripts indicate derivatives with respect to the variables. The~direction of the $x$-axes follows the sketch in Figure~\ref{sketch}.

At the meeting point between the three branches, i.e., $x=0$, we have the boundary \mbox{conditions~\cite{grun93,koga00}}
\begin{equation}
u^{(1)}=u^{(2)}+u^{(3)}, \qquad \qquad u^{(1)}_x=u^{(2)}_x=u^{(3)}_x.
\label{bp}
\end{equation}

The first equation comes from the physical property that the magnetic flux through an infinitesimally small
contour encircling the origin must vanish, i.e., the total change of the gauge-invariant phase difference is zero. The~second equation means that the field, which is proportional to the slope of the phase difference, is continuous at the origin. 

Far away from the origin $x\ll0$, the sG Equation \eqref{sg} admits two types of fundamental solitons: a kink, that is in the form of
\begin{equation}
u(x,t) = 4 \tan^{-1}\left\{\exp[-\gamma(x - vt - x_0)] \right\}, \qquad x_0 \in \mathbb{R}
\label{kink}
\end{equation}
and a breather
\begin{equation}
u(x,t) = 4\tan^{-1}\left\{\tan\theta\sin\left[\gamma\cos(\theta)(t-vx-t_0))\right]\sech\left[\gamma\sin(\theta)(x-vt-x_0)\right]\right\}, \qquad x_0 \in \mathbb{R}
\label{breather}
\end{equation}
where $\gamma=1/\sqrt{1-v^2}$ is the Lorentz-type contraction of a moving excitation. Both~represent a topological and non-topological soliton moving with velocity $v$. The~breather \eqref{breather} oscillates in time with frequency $\gamma\cos\theta$. All the previous work on solitons in tricrystal junctions dealt with the first class of solutions (see the related references mentioned in Section \ref{sec2}). Here, we will concentrate on breather~dynamics. 

To study the dynamics of small-amplitude breathers~\eqref{breather}, one can also use the standard multiple-scale expansion method. Writing 
\[
u^{(j)}(x,t) = \epsilon U^{(j)}(X,T) + \text{c.c.} + \text{higher-order terms},
\]
where $X=\epsilon x$ and $T=\epsilon^2t/2$ are the slow space and time variables, respectively, and c.c. is the complex conjugation of the preceding terms, we obtain the NLS equation (see, e.g.,~\cite{kram13})
\vspace{12pt}
\begin{eqnarray}
iU^{(j)}_T+U^{(j)}_{XX}+\frac12|U^{(j)}|^2U^{(j)}=0.
\label{nls}
\end{eqnarray}

The boundary conditions \eqref{bp} then become 
\begin{equation}
U^{(1)}=U^{(2)}+U^{(3)}, \qquad \qquad U^{(1)}_X=U^{(2)}_X=U^{(3)}_X.
\label{bp2}
\end{equation}

Far away from the branch point $X=0$, i.e., $X\to-\infty$, the NLS equation has a traveling bright~soliton 
\begin{equation}
U(X,T)=A\sech(A(X-vT-X_0)/2)\exp(i\phi-ivX/2+i(A^2 -v^2)T/4), 
\label{bs}
\end{equation}
with $A>0$, $X_0$, $\phi$ and $v\in\mathbb{R}$. This soliton approximates the breather solution \eqref{breather} for small $|\theta|$.

\section{ Scattering of NLS Solitons}		\label{sec3}

Studying the scattering problem for non-topological solitary waves as sketched in Figure~\ref{sketch}, we can consider a symmetric solution between the second and third branch, i.e., $u^{(2)}=u^{(3)}$ and $U^{(2)}=U^{(3)}$. Under the symmetry, the initial boundary value problems \eqref{sg}--\eqref{bp} and \eqref{nls}--\eqref{bp2} become a single equation on the real line:
\vspace{-12pt}
\begin{eqnarray}
u_{xx}-u_{tt}=\sin u, \qquad x\in\mathbb{R},\label{sg1}
\end{eqnarray}
{with} $u(0^-)=2u(0^+),\:u_x(0^-)=u_x(0^+)$ and
\begin{eqnarray}
iU_T+U_{XX}+\frac12|U|^2U=0, \qquad X\in\mathbb{R},\label{nls1}
\end{eqnarray}
with $U(0^-)=2U(0^+),\:U_X(0^-)=U_X(0^+)$. 

\subsection{Scattering in the Linear Problems}

It is natural to consider first the scattering problem for plane waves in the linear Schr\"odinger equation obtained by omitting the cubic term in Equtaion~\eqref{nls1}. Incoming plane waves arriving from $X \to -\infty$ has the form $U(X,T) = e^{i(kX-\omega T)}$, with the amplitude normalized to 1, $\omega > 0$, and $k=\sqrt\omega$.  The~general solution of the scattering problem is
\begin{eqnarray}
\psi(X)=\left\{
\begin{array}{llll}
e^{i\left(kX-\omega T\right)}+\tilde{r}e^{i\left(-kX-\omega T\right)}, \qquad X < 0,\\
\tilde{t}e^{i\left(kX-\omega T\right)}, \qquad X > 0.
\end{array}
\right.
\label{scat}
\end{eqnarray}

Here, $\tilde{r}$ and $\tilde{t}$ are reflection and transmission coefficients, respectively. 

Substituting the solution \eqref{scat} into the boundary conditions at $X=0$ will give us 
\begin{eqnarray}
\tilde{r} = \frac13, \qquad \tilde{t}= \frac23.
\label{linsc}
\end{eqnarray}

It turns out that these coefficients do not satisfy the standard (unitarity) condition due to the~following:
\begin{equation*}
|\tilde{r}|^2+|\tilde{t}|^2 = \frac{5}{9} < 1, \qquad \tilde{t} < 1 + \tilde{r} = \frac{4}{3}.
\end{equation*}

This informs us that the vertex conditions will not preserve the `mass' of localized solutions $\mathcal{M}=\int_{X\in\mathbb{R}} |U|^2\,dX$. 

\subsection{NLS Soliton Scattering}

We now consider the interaction of a bright soliton with the vertex. We integrate the NLS equation~\eqref{nls1} numerically using the fourth-order Runge-Kutta method. The~Laplacian is discretized using a three-point central difference. Therefore, our scheme has the discretization error of at least order $\mathcal{O}(\Delta x^2,\Delta t^4)$, where $\Delta x$ and $\Delta t$ are the spatial mesh size and the time step, respectively. We used several combinations of  $\Delta x$ and $\Delta t$ to make sure that variations of the computed results caused by the discretization are small enough. As the initial data, we take a soliton approaching the branch point from $X\ll0$ in the form of $U(X,0)$ from \eqref{bs}, and without losing its generality, we set $A = 1$.

\begin{figure}[htbp]
\centering
\subfigure[]{\includegraphics[scale=0.39,clip=]{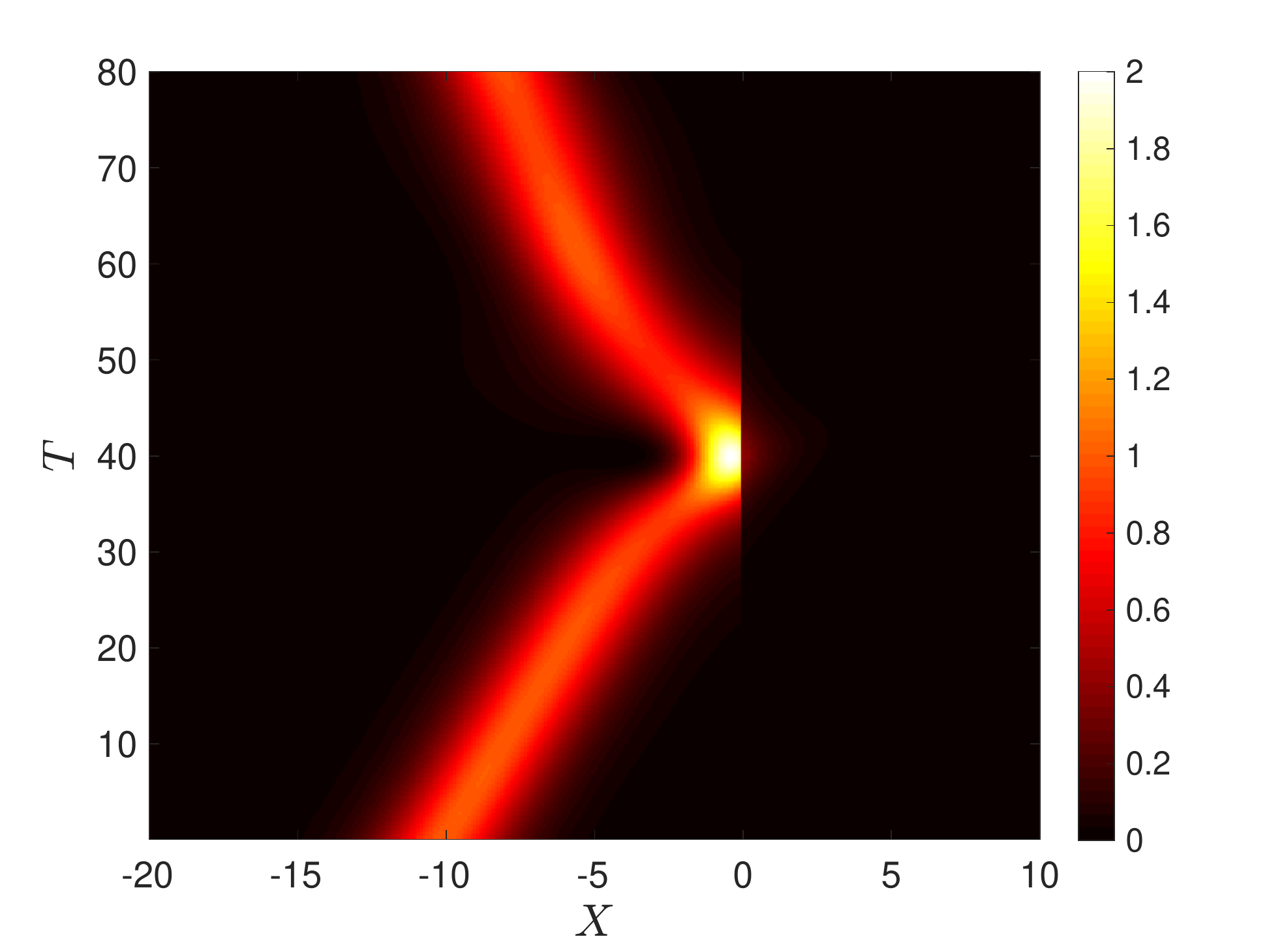}\label{tda}} 
\subfigure[]{\includegraphics[scale=0.39,clip=]{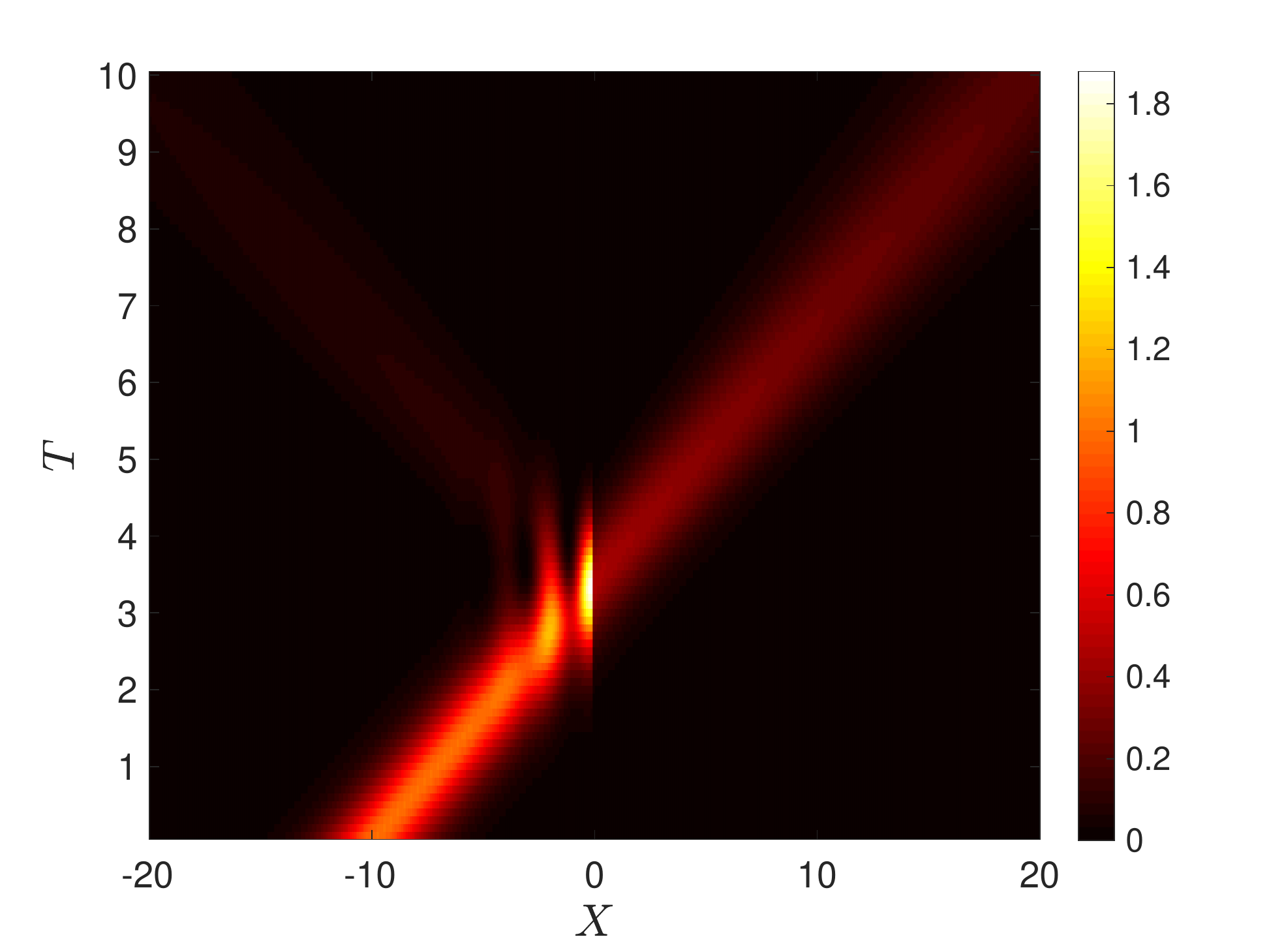}\label{tdb}}\\
\subfigure[]{\includegraphics[scale=0.5]{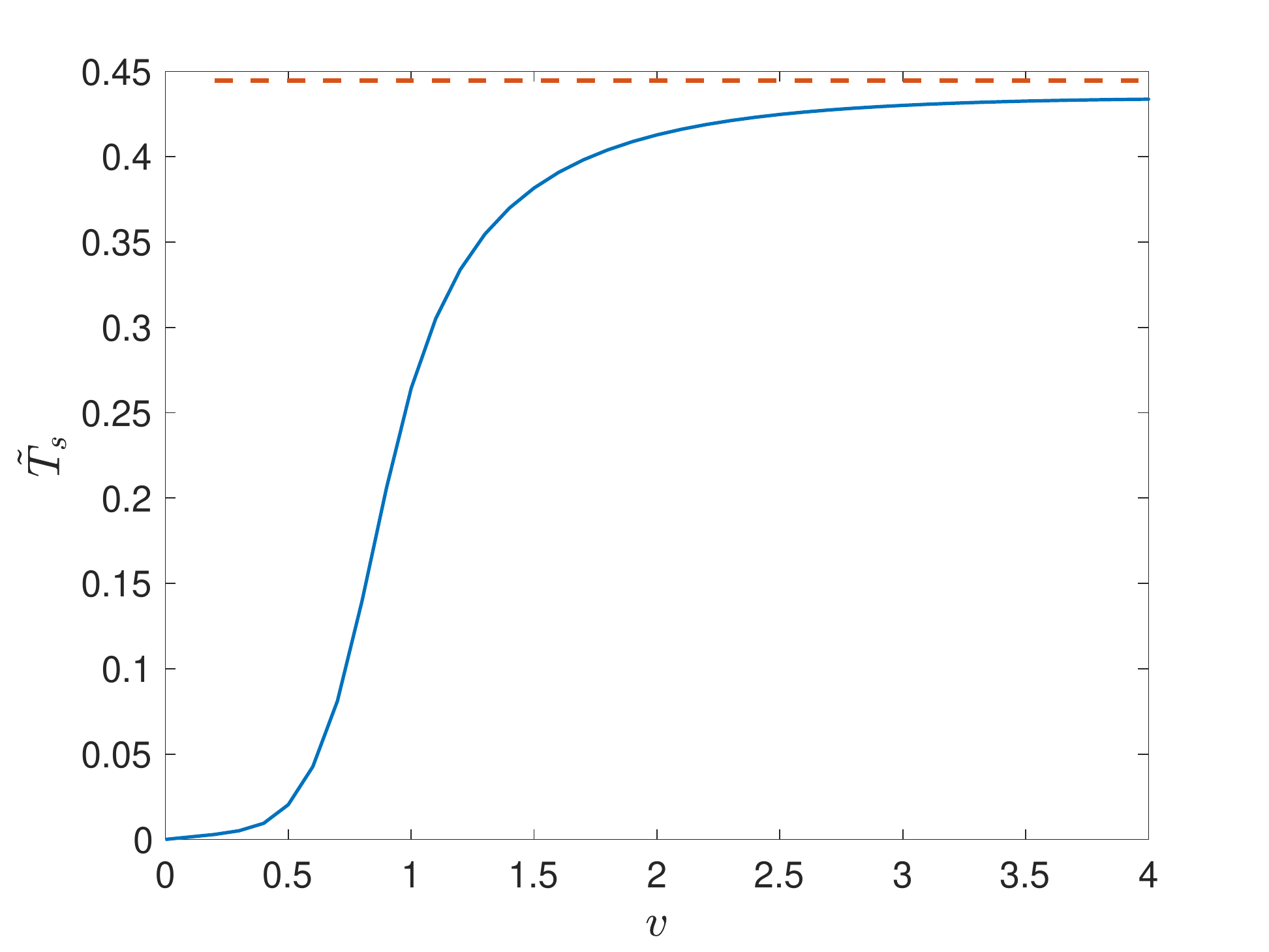}\label{tdc}}
\caption{(\textbf{a},\textbf{b}) Dynamics of a soliton moving towards the branch point $X=0$ with the initial velocity $v=0.3$ (\textbf{a}) and $v=3$ (\textbf{b}). Shown are the top view of $|U(X,T)|^2$. (\textbf{c}) Numerically obtained transmission rate $\tilde{T}_s$ as a function of the incoming soliton velocity $v$. The~horizontal dashed line is the theoretical approximation $\tilde{t}^2 = \frac{4}{9}$ from \eqref{linsc}.} \label{td}
\end{figure}

In Figure~\ref{td} we plot the dynamics of a soliton traveling towards the origin for two different initial velocities, namely $v=0.3$ and $3$, representing slow- and fast-moving solitons, respectively. In~both cases, especially for the slow-incoming soliton, we see that as it approaches the branch point, it accelerates. This is usually a characteristic of an attractive potential. However, note that after the interaction, there is no trapped state, which on the other hand is a characteristic of a repulsive potential. It can be easily checked that the  corresponding linear eigenvalue problem of \eqref{nls1} has no point spectrum, which confirms the absence of a trapped state. Thus, our branching point has both characteristics of an attractive as well as a repulsive potential at the same time.
 
The main difference between Figure~\ref{td}a,~b is that for the slow-incoming soliton, most of the mass is reflected, while for the fast-moving one, most of it is transmitted. To quantify how much of the mass is being transmitted, we define the `transmission rate' $\tilde{T}_s$ as
\begin{equation}
\tilde{T}_s = \lim_{T\to\infty} \frac{1}{4A} \int_{X>0}  |U(X,T)|^2 \, dX.
\label{tts}
\end{equation}

Here, we normalize the rate with the initial $L^2$-norm $\lim_{T\to-\infty}||U(T)||_{L^2}^2 = 4A$, which (in the absence of the vertex conditions) normally would be constant in time. We plot the transmission rate $\tilde{T}_s$ in Figure~\ref{td}c for different values of the initial velocity $v$. There is a steep transition in the interval of $v\approx0.5$ and $v\approx1.5$ where the soliton changes from being mostly reflected to mostly transmitted. 

Recalling that for large $v$, the essential wave numbers of the Fourier transform of NLS solitons are concentrated around $k = v$, we can expect that the quantum transmission rate of a soliton with velocity $v$ will approach a limiting value that is given by the absolute value square of the transmission coefficient of the linear plane wave $|\tilde{t}|^2$, see \eqref{linsc}. We can see that this is indeed the case.

\section{Scattering of sG Breathers}		\label{sec4}
After studying the soliton scattering in the NLS setting, finally, we now consider the original problem, i.e., scattering in the sG equation context. The~results of Section \ref{sec3} should be comparable to the scattering of small-amplitude breathers in the sG equation.

In the infinite domain problem without any vertex condition at $x=0$, the sG equation preserves the `mass' $\mathcal{H}=\int_{x\in\mathbb{R}} H\,dx$, where the function $H$ is the Hamiltonian given by 
\begin{equation}
H(x,t) = \frac12u_t^2+\frac12u_x^2+(1-\cos u).
\end{equation}

We solved the sG Equation \eqref{sg1} using a similar numerical integration method previously implemented in solving the NLS equation in Section \ref{sec3}. As the initial data, we take the breather~\eqref{breather} with $x_0\ll0$ at $t=0$. 

\begin{figure}[H]
\centering
\subfigure[]{\includegraphics[scale=0.39,clip=]{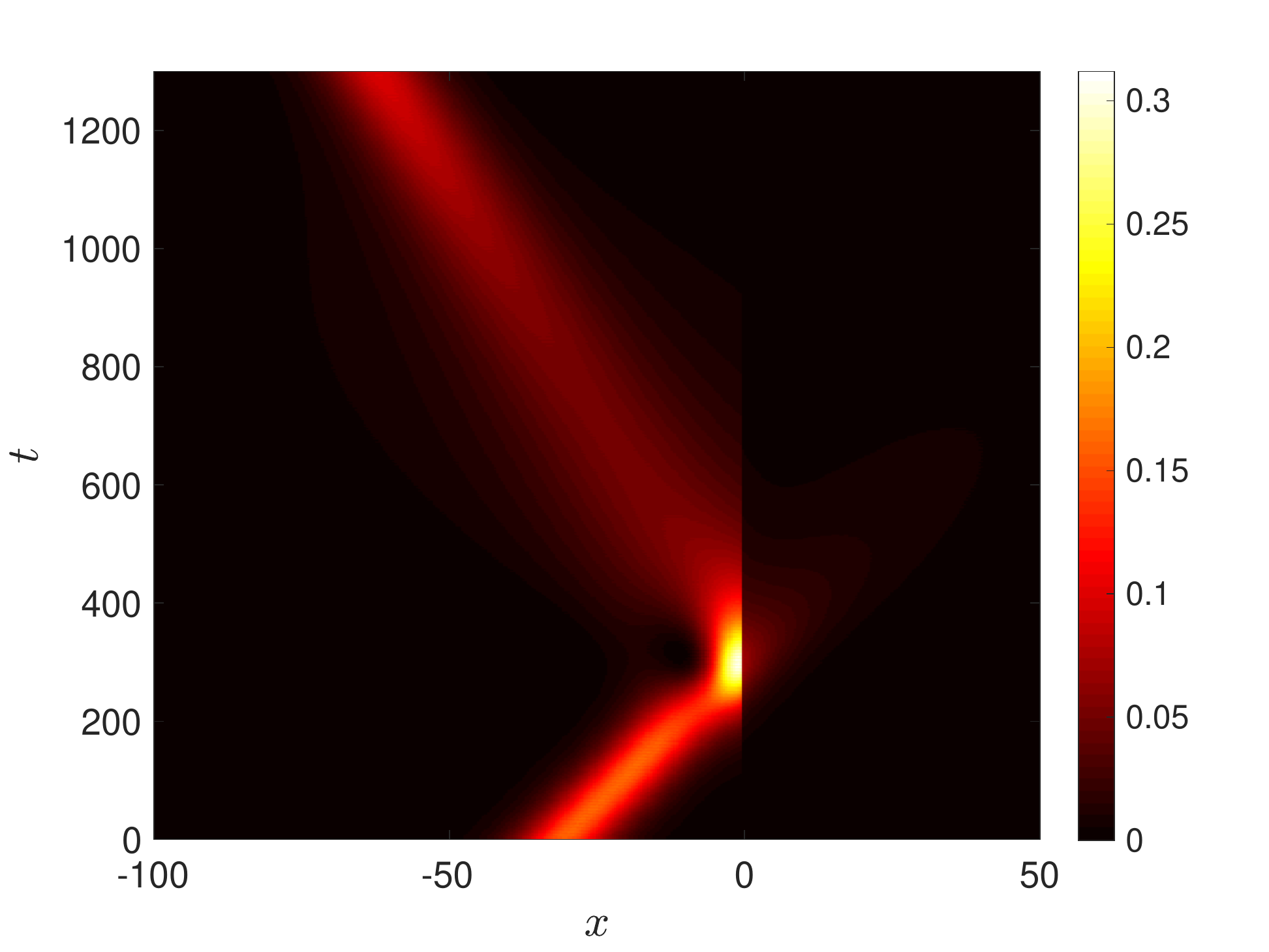}\label{sga}} 
\subfigure[]{\includegraphics[scale=0.39,clip=]{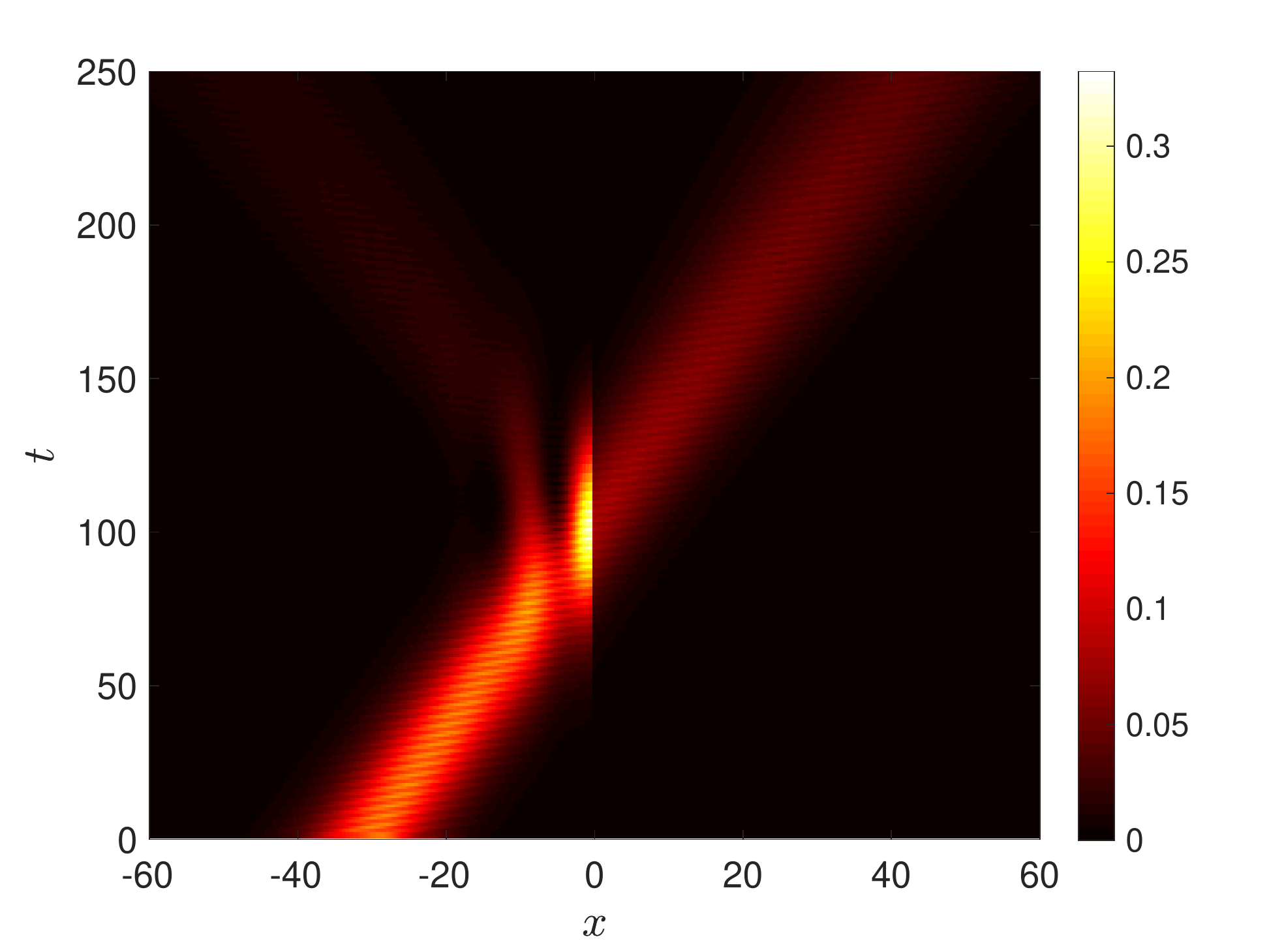}\label{sgb}}\\ 
\subfigure[]{\includegraphics[scale=0.5]{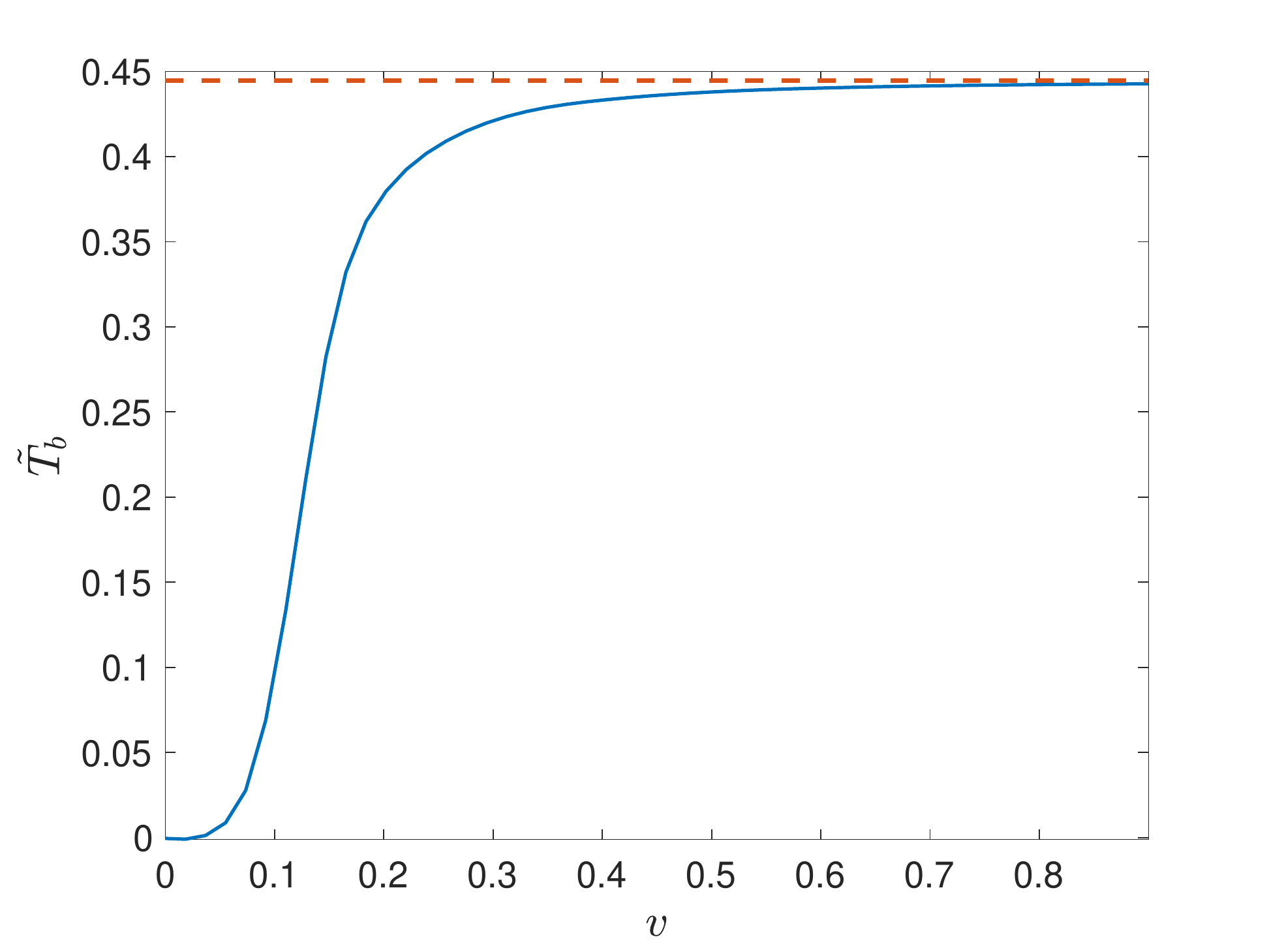}\label{trsg}}
\caption{Similar to Figure~\ref{td}, but for (small-amplitude) sG breathers with velocity $v=0.1$ (\textbf{a}) and $v=0.3$ (\textbf{b}). In~panels (\textbf{a},\textbf{b}), shown are the top view of $H(x,t)$. Panel (\textbf{c}) is the transmission rate $\tilde{T}_b$ as a function of the incoming breather velocity $v$. The~horizontal dashed line is the theoretical approximation $\tilde{t}^2 = \frac{4}{9}$ from \eqref{linsc}. Here, $\theta=\cos^{-1}0.99$.}  \label{tsg}
\end{figure}

First, we simulate the dynamics of small-amplitude breathers with $\theta=\cos^{-1}0.99$. In~Figure~\ref{tsg}, we present our simulations of a breather traveling with two different velocities towards the branch point $x=0$. It is interesting to see a close resemblance between panels (a, b) of Figure~\ref{tsg} and those of Figure~\ref{td}. It is then instructive to compute the transmission rate of the breathers for different values of the incoming velocity. Defining
\begin{equation}
\tilde{T}_b=\lim_{T\to\infty} \frac{1}{\mathcal{H}_0}\int_{x>0}H(x,t)\,dx,
\label{tts1}
\end{equation}
where $\mathcal{H}_0=\lim_{t\to-\infty}||H(t)||_{L^1}$, we plot in Figure~\ref{tsg}c the transmission rate $\tilde{T}_b$, which again compared to Figure~\ref{td}c shows the same qualitative profile. Moreover, we obtain numerically that the nonlinear transmission rate tends to the linear one $\tilde{t}^2$ \eqref{linsc} as the incoming breather velocity $v \rightarrow 1$.

\begin{figure}[H]
\centering
\subfigure[]{\includegraphics[scale=0.39,clip=]{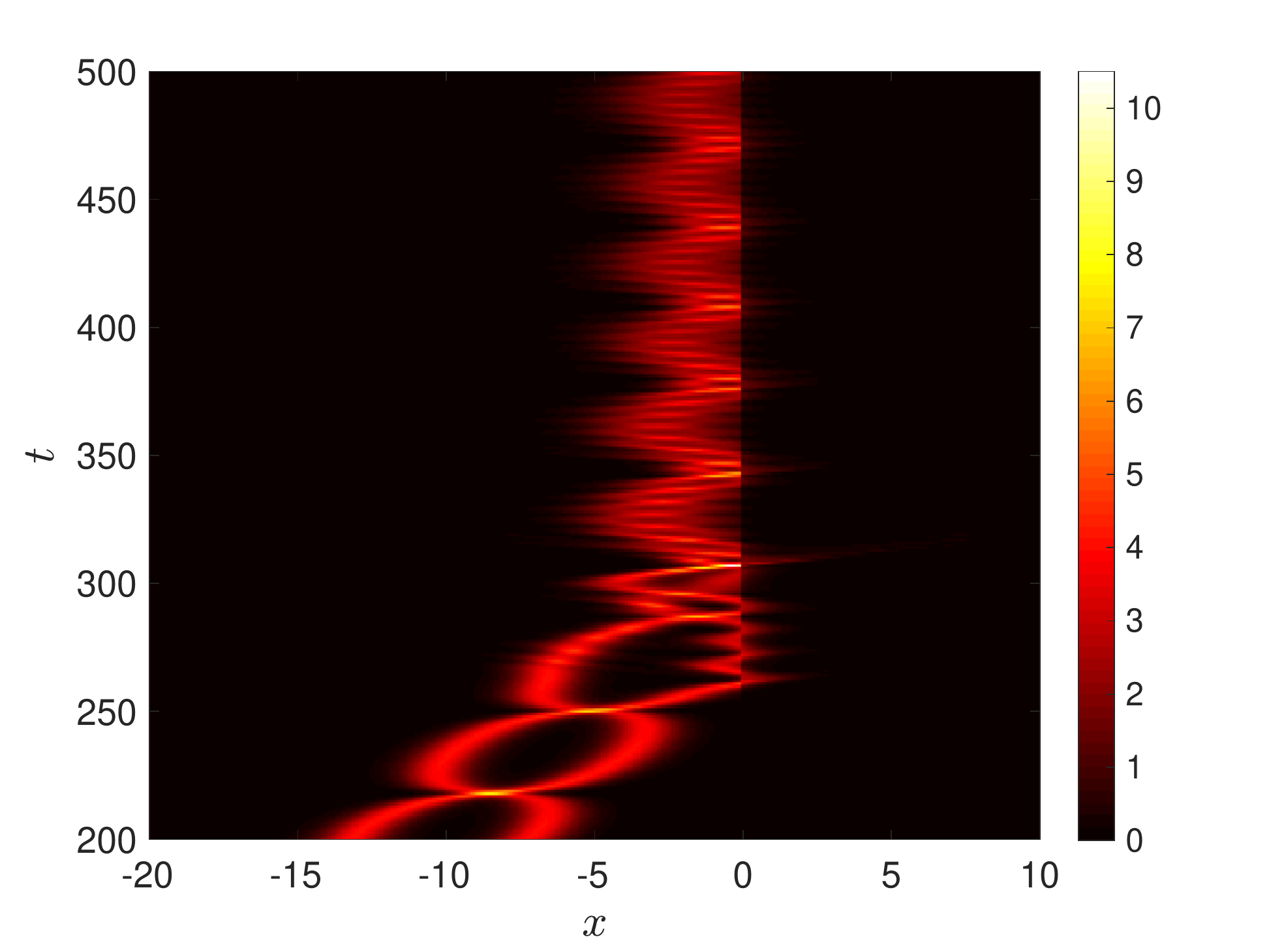}\label{sga2}}
\subfigure[]{\includegraphics[scale=0.39,clip=]{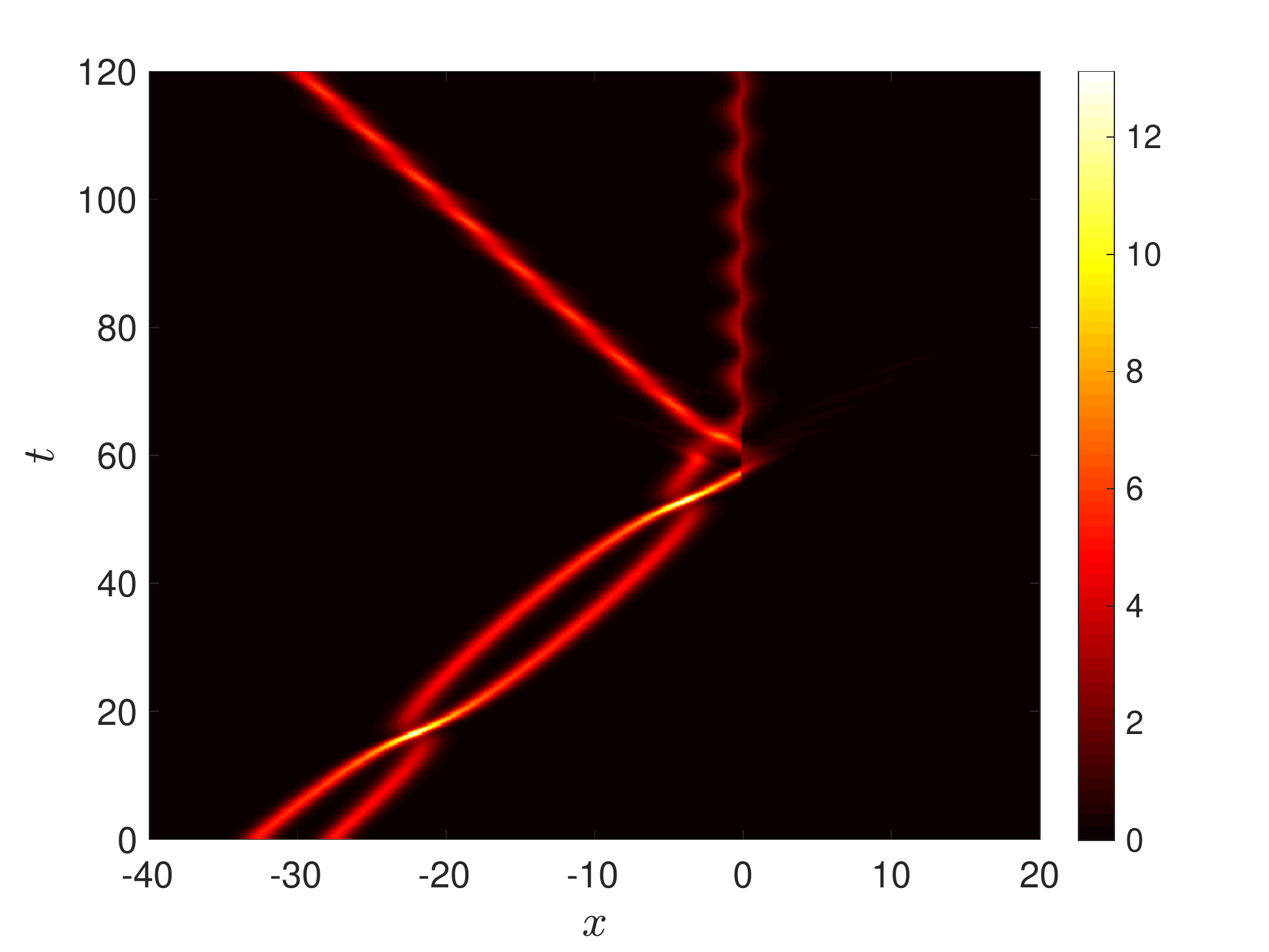}\label{sgb2}}
\caption{Similar to Figure~\ref{tsg}a,b, but for a large-amplitude sG breather with velocity $v=0.1$ (\textbf{a}) and $v=0.5$ (\textbf{b}). Here, $\theta=\cos^{-1}0.1$.}	\label{tsg2}
\end{figure}

Next, we consider large-amplitude breathers. In~a similar setup, we simulated a slow- and a fast-incoming soliton. We present the typical dynamics of the two cases in Figure~\ref{tsg2}, where we take $\theta=\cos^{-1}0.1$. 

Far away from the branch point, a large-amplitude breather can be seen obviously as an oscillating pair of a kink and an anti-kink. Upon collision with the vertex, the slow-moving breather is trapped at the origin, while the fast-moving one dissociates into a trapped kink and an ejected anti-kink. Both cases show completely different dynamics from the small-amplitude breathers and the NLS solitons, see Figures \ref{td} and \ref{tsg}. From the simulations, we note that the vertex acts as a repulsive potential for small-amplitude breathers, while it behaves as an attractive potential for large-amplitude ones. The~difference can be explained as follows. 

The branch point has been reported before to trap kinks, i.e., topological excitations~\cite{naka76,grun93,hatt96,susa05}. They are given analytically by~\cite{grun93,susa05}
\begin{equation}
u=4\tan^{-1}\left\{\exp[-x\pm\ln\sqrt3]\right\},
\end{equation}
with the `$+$' sign for the region $x<0$ and the `$-$' sign for $x>0$. They also have an oscillatory mode with frequency~\cite{grun93,susa05}
\begin{equation}
\omega=\sqrt{\frac{1+\sqrt{13}}{8}}.
\label{w1}
\end{equation}

When excited, the mode will vibrate, but eventually, it will fade away because of 'radiative' damping, i.e., a damping mechanism due to the excitation of higher harmonics with frequencies in the continuous spectrum. Even more, trapped kink oscillations decay asymptotically at a rate of $\mathcal{O}(t^{-1/2})$, see~\cite{ali11}. Here, even though our excitations are non-topological, large-amplitude breathers are close to an intertwining pair of kink and anti-kink. Trapping is therefore expected and in this case, an oscillatory mode must exist. Such oscillation can be seen quite well in Figure~\ref{tsg2}b where the trapped kink jiggles about $x=0$. 

As for the oscillatory dynamics of the breather in Figure~\ref{tsg2}a, we can explain it as follows. The~breather does not dissociate into a separate kink and anti-kink because its free energy is not enough to do so. As a result, it maintains its non-topological shape. In~return, the vertex will tend to repel it. However, the non-topological excitation still has a relatively large amplitude, which on the other side looks like a kink and hence tends to be attracted by the vertex. Therefore, the branch point acts as a repellent and an attractor at the same time. This creates the oscillatory movement of the breather. Additionally, we also observe a fast oscillation. We strongly suspect that it is due to the trapping mode and hence its frequency is approximately given by \eqref{w1}. 

\section{Conclusions}
\label{sec5}

We have analyzed for the first time the dynamics of breathers in a tricrystal Josephson junction. The~physically relevant model consists of three semi-infinite Josephson junctions coupled at one end. For small-amplitude breathers, we have derived the corresponding NLS equation on star graphs from the governing sG equation. We have shown the resemblance of the dynamics of small-amplitude breathers with the NLS bright solitons. Large-amplitude breathers yield qualitatively different~behaviors. 

For future work, it will be interesting to rigorously study the collision of a fast-incoming NLS soliton with the vertex. This study will be along the lines of the analysis by Holmer et al.~\cite{holm07} about the scattering of fast-moving solitons by a delta interaction on the line. An extension of their work to the case of sG equations on graphs, which is not available yet, will also be particularly appealing. In~the context of collective coordinate methods, it will be important to derive an effective Hamiltonian describing slow-moving soliton interactions with the vertex, along the idea of, e.g.,~\cite{fori94,good04,holm07a}. The~main challenge is how to incorporate the vertex condition, which is not explicitly embedded within the governing equation. On the numerical side, it will also be interesting to provide bounds to the discretization as well as the round-off errors of our numerical scheme.

\vspace{6pt} 

\authorcontributions{The contributions of the respective authors are as follows: H.S.~designed and  supervised the research and wrote the article; N.K.~conducted analytical computations and wrote the article; Z.Z.~performed numerical computations; T.N.~and T.W.~prepared the initial work and revised the article. All authors read and approved the final manuscript.}

\funding{N.K.\ was funded by the National Research Foundation (NRF) of Korea through Grant No. NRF-2017-R1C1B5-017743 under the Basic Research Program in Science and Engineering.}

\acknowledgments{The authors acknowledge the two referees for their comments that improved the paper and Andrew Harrison and Christopher Saker (University of Essex) for proofreading the manuscript.}

\conflictsofinterest{The authors declare no conflict of interest.} 

\abbreviations{The following abbreviations are used in this manuscript:\\

\noindent 
\begin{tabular}{@{}ll}
NLS & nonlinear Schr\"odinger\\
sG & sine-Gordon\
\end{tabular}}


\reftitle{References}

\end{document}